\documentclass[a4paper,12pt]{article}
\usepackage{tikz}
\usepackage{graphicx}

\usepackage{mathtools}
\usepackage{braket}
\usepackage{xcolor}
\usepackage{ulem}
\usepackage[utf8]{inputenc}
\usepackage[english]{babel}
\usepackage{amsmath,amsfonts,amssymb}
\usepackage{mathpazo} 
\usepackage[scaled]{helvet}
\usepackage{url}
\usepackage[colorlinks=true, allcolors=blue]{hyperref}
\usepackage{authblk}
\usepackage{xifthen}
\usepackage{textcomp} 
\usepackage{graphicx,xcolor}
\usepackage{wrapfig} 
\usepackage{colortbl}
\usepackage{booktabs}
\usepackage{algorithm}
\usepackage[noend]{algpseudocode}
\usepackage{changepage}
\usepackage[numbers,sort&compress]{natbib}
\usepackage{fancyhdr}  
\usepackage{lastpage}  
\usepackage[explicit]{titlesec}
\usepackage{enumitem} 
\usepackage[resetlabels]{multibib}
\usepackage{letltxmacro}
\usepackage{xparse}
\usepackage{silence}
\usepackage{xpatch}

\title{Photonics in highly dispersive media: The exact modal expansion}

\author{Frédéric Zolla, André Nicolet and Guillaume Demésy\footnote{Corresponding author : \texttt{guillaume.demesy@fresnel.fr}}\\
\footnotesize{Aix Marseille Univ, CNRS, Centrale Marseille, Institut Fresnel, Marseille, France.}
}

\def\C{\mathbb C}
\def\Curl{\mathrm{curl}}
\def\TE{\mathrm{T}_{\varepsilon,\mu}}

\def\TET{\mathrm{T}_{\varepsilon^T,\mu^T}}
\def\lambdas{\lambda_s}
\def\omegas{\omega_s}
\def\KOmega{\mathcal{D}}
\def\KU{{\mathcal{U}}}

\begin{document}
\maketitle

\begin{abstract}
We present exact modal expansions for photonic systems including highly dispersive media. The formulas, based on a simple version of the Keldy\v{s}
theorem, are very general since both permeability and permittivity can be dispersive, anisotropic, and even possibly non reciprocal. A simple
dispersive test case where both plasmonic and geometrical resonances strongly interact exemplifies the numerical efficiency of our approach.
\end{abstract}

\section{Introduction}
Recently, modal expansion formalisms have received a lot of attention in photonics because of their capabilities to model the physical properties in the natural resonance-state basis of a considered system, leading to a transparent interpretation of the numerical results\cite{bai2013efficient,sauvan2013theory,vial2014quasimodal,yan2018rigorous}. Their practical use requires to take into account time dispersive media and this leads to non Hermitian (i.e. with complex eigenfrequencies) non linear eigenvalue problems \cite{spence2005photonic,engstrom2017rational}. Nowadays, efficient numerical algorithms are available for the numerical computations of such problems \cite{Hernandez:2005:SSF,voss2013nonlinear,guttel2017nonlinear} but the use of these nonlinear eigenmodes in modal expansions is still an active research topic. In this paper, we propose a straightforward approach based on the Keldy\v{s} theorem \cite{keldysh1951eigenvalues,keldysh1971completeness,kozlov1999differential,beyn2012integral,van2016nonlinear,unger2013convergence} that provides modal expansions for very general systems (both permittivity and permeability are dispersive, anisotropic, and possibly non reciprocal). In this approach, the central concept is the set of the eigentriplets associated to the non linear eigenproblem and no normalization or biorthogonality are involved. In the simpler cases, the expansion is similar to recent results for dispersive permittivity\cite{leung1994completeness,ge2014quasinormal,perrin2016eigen,muljarov2016resonant,doost2016resonant,lalanne2018light,yan2018rigorous}. As a proof of concept, we provide the example of a simple 2D cavity loaded with a piece of dispersive Lorentz medium.


\section{The dispersive modal expansion}

\subsection{The non linear eigenvalue problem and the Keldi\v{s} theorem}
We consider holomorphic operators $T(z)$, that is a set of (possibly non self-adjoint) operators parametrized by a complex parameter $z$ \cite{kozlov1999differential} 
and the associated eigenvalue problem:
given a holomorphic operator $T(z)$, we define its \textit{eigentriplets} as ordered sets $(\lambda ,\bra{\mathbf{u_l}}, \ket{\mathbf{u_r}})$ of an eigenvalue and its associated \textit{left} and \textit{right} eigenvectors satisfying $\bra{\mathbf{u_l}} T(\lambda)=0$ and $T(\lambda) \ket{\mathbf{u_r}} =0$. An eigenvalue is simple if $T'(\lambda) \ket{\mathbf{u_r}}  \neq 0$ where $T'(z)$ is the complex derivative of $T(z)$ and, in the sequel, we will consider that all the eigenvalues are simple or at least semi-simple 
and that the operators are diagonalizable. In practice, finding the couples $(\lambda ,\bra{\mathbf{u_l}})$ satisfying $\bra{\mathbf{u_l}} T(\lambda)=0$ amounts to looking for \textit{right} eigenvectors $\ket{\mathbf{u_l}}$ such that $T^*(\lambda) \ket{\mathbf{u_l}} =0$ where $T^*$ is the adjoint operator associated with $T$ and, in the sequel, $\bra{\mathbf{u_l}} \ne \bra{\mathbf{u_r}}$ due to the non self-adjointness of the operators at stake.

The fundamental result that we are using is the Keldy\v{s} theorem \cite{keldysh1951eigenvalues,keldysh1971completeness}. For the sake of
simplicity, we give here its version for matrices with simple eigenvalues as stated by Van Barel and Kravanja\cite{van2016nonlinear} in their
presentation of Beyn's algorithm \cite{beyn2012integral} (an integral method for solving nonlinear eigenvalue problems):
\textbf{Theorem} (Keldy\v{s}):
Given a domain $\KOmega\subset\C$ and an integer $m \geq 1$, let $\mathcal{C} \subset \KOmega$ be a compact subset, let $T$ be a matrix-valued function $T : \KOmega \longrightarrow \C^{m \times m}$ analytic in $\KOmega$
and let $n(\mathcal{C})$ denote the number of eigenvalues of $T$ in $\mathcal{C}$.
Let $ \lambda_k$ for $k = 1, \ldots, n(\mathcal{C})$  denote these eigenvalues and suppose that all of them are simple. Let $\ket{\mathbf{u_l}_{k}}$ and $\bra{\mathbf{u_r}_k}$ for $k = 1, \ldots, n(\mathcal{C})$ denote their left and right eigenvectors, such that
\[
T(\lambda_k) \ket{\mathbf{u_r}_{k}} =0 ,  \; \; \; \; \; \; \; \; \;    \bra{\mathbf{u_l}_k} T(\lambda_k)=0.
\]
Then there is a neighborhood $\KU$ of $\mathcal{C}$ in  $\KOmega$ and a matrix-valued analytic function $R : \KU \longrightarrow \C^{m \times m}$ such that the resolvent $T(z)^{-1}$ can be written as
\begin{equation}
T(z)^{-1}=  \sum_{k=1}^{n(\mathcal{C})} \frac{1}{(z-\lambda_k)} \frac{\ket{\mathbf{u_r}_{k}} \,   \bra{\mathbf{u_l}_k}}{\braket{\mathbf{u_l}_k |T'(\lambda_k) \mathbf{u_r}_k } }+ R(z) \label{keldysh}
\end{equation}
for all $z \in \KU \setminus \{\lambda_1 ,\ldots, \lambda_{n(\mathcal{C})} \}$.
If $T$ is a matrix-valued strictly proper rational function, the analytic function $R$ is equal to zero \cite{van2016nonlinear}.
Considering the similar results for holomorphic operator functions does not cause additional difficulties \cite{kozlov1999differential}. We work here with the differential operators of electrodynamics but, for all practical purposes, the numerical computations are performed using matrices.

\subsection{Modal expansion for electrodynamics in the presence of time-dispersive materials}

In this section, we apply the Keldi\v{s} theorem to the modal expansion in electrodynamics \textit{i.e.} using the electromagnetic eigenmodes computed using some nonlinear eigenvalue numerical solver such as SLEPc\cite{Hernandez:2005:SSF} for fast computation of direct scattering problems.
We start with the Maxwell's equations in harmonic regime with a given source represented for instance by a density of current $\mathbf{j}$. We then have to find, in some $\Omega \subset \mathbb{R}^3$ with given boundary conditions, the electromagnetic field $(\mathbf{E_s},\mathbf{H_s})$ such that $ \Curl \mathbf{E_s} = -\lambdas \mu(\lambdas) \mathbf{H_s}$ and $\Curl \mathbf{H_s} = \lambdas \varepsilon(\lambdas) \mathbf{E_s} + \mathbf{j}$  where $\lambdas$ is a pure imaginary number given by $\lambdas :=i \omegas$ and where $\varepsilon(\lambdas)$ and $\mu(\lambdas)$ are the time-dispersive permittivity and permeability respectively which are then taken as functions of $i\omegas$. We do not put special restrictions on these permittivity and permeability: they can be anisotropic and non-reciprocal, they are therefore general tensor densities represented by general invertible non Hermitian matrices (in fact invertible matrix-valued functions of $\lambda \in \C$ except for a finite set of points). In practice, the time-dispersion can be very accurately represented by rational functions \cite{Garcia-Vergara:17}.
Classically, one of the field is eliminated to obtain a wave equation and we start with the electric formulation:
\[
\Curl \, ( \mu(\lambdas)^{-1} \Curl \, \mathbf{E_s}) + \lambdas^2 \varepsilon(\lambdas) \, \mathbf{E_s} = -\lambdas \mathbf{j} \, .
\]
The problem is then posed as a problem of the inversion on the following operator: $\TE(\lambdas) := \Curl \, ( \mu(\lambdas)^{-1} \Curl \, \cdot) + \lambdas^2 \varepsilon(\lambdas) \cdot \,$.
The ket $\ket{\mathbf{F_r}}$ is then represented by a vector field $\mathbf{F_r}$, the bra $\bra{\mathbf{E_l}}$ by another one, $\mathbf{E_l}^*:=\overline{\mathbf{E_l}}^T$, and the braket product $\braket{\mathbf{E_l}|\mathbf{F_r}}= \int_{\Omega} \overline{\mathbf{E_l}} \cdot \mathbf{F_r} \, d \Omega$, using the dot product, 
or equivalently $\braket{\mathbf{E_l}|\mathbf{F_r}}= \int_{\Omega} \mathbf{E_l}^* \mathbf{F_r} \, d \Omega$, using the matrix product.
The Helmholtz equation can be summed up in the following fashion: $\TE(\lambdas) \ket{\mathbf{E_s}}=  -\lambdas \ket{\mathbf{j}}$. Our purpose is to express $\ket{\mathbf{E_s}}$ as an expansion of the right eigenvectors associated with the 
eigenproblem $\TE(\lambda) \ket{\mathbf{E}}=0$.
In order to apply (\ref{keldysh}), we explicitly need $\TE'(\lambda)$.
Our final purpose is to use the formula in numerical computations and this imply the discretization of the problem. This is performed via a weak formulation and a finite element process. It has been shown that the edge elements and their later generalizations are the most appropriate to avoid spurious modes in the electrodynamic eigenmode computations\cite{bossavit1990spurious}. We will perform the computations with the Maxwell differential operators and use an integration by part to produce a practical formula that can be transposed directly to the discrete fields and the matrices.

 It remains to explicitly compute the adjoint $\TE^*(\lambda)$.
For that we use
$\int_\Omega \overline{\mathbf{u}} \cdot \Curl \mathbf{v} \, d\Omega =  \int_\Omega \Curl \overline{\mathbf{u}}  \cdot\mathbf{v} \, d\Omega +  \int_{\partial \Omega} (\overline{\mathbf{u}}\times\mathbf{v}) \cdot\mathbf{n} dS$. By integrating twice by parts the braket $\braket{\mathbf{E_l} | \TE(\lambda) \mathbf{F_r}} $, we obtain:
\[
\int_\Omega \overline{\mathbf{E_l}} \cdot \left( \TE(\lambda) \mathbf{F_r} \right)\, d\Omega  = \int_\Omega \left( \TET(\lambda)  \overline{\mathbf{E_l}} \right)\cdot  \mathbf{F_r} \, d\Omega  +b.t.
\]
where $b.t.$ stands for the boundary terms appearing in the integration by parts and where $\TET(\lambda) = \Curl \, \mu(\lambda)^{-T} \Curl \cdot + \lambda^2 \, \varepsilon(\lambda)^T \cdot$ (the superscript $^{-T}$ standing for transposition of the inverse matrix). In order to find the adjoint, the previous expression has to be conjugated which needs to be cautiously done. $\TE^*(\lambda)=\Curl \, \overline{\mu}(\lambda)^{-T} \Curl \cdot + \overline{\lambda^2} \, \overline{\varepsilon}(\lambda)^T \cdot$. But due to the Hermitian symmetry of the permittivity and the permeability, we have $\overline{\varepsilon}(\lambda)=\varepsilon(\overline{\lambda})$ and $\overline{\mu}(\lambda)=\mu(\overline{\lambda})$, {\textit{i.e.}}, $\TE^*(\lambda)=\TET(\overline{\lambda})$.
 We can observe that in the case of isotropic media represented by scalar parameters and of reciprocal media represented by symmetric tensors,
the transposition is an identity and only the complex conjugation remains.
 The eigenvector $\mathbf{E_l}$ may therefore be simply the complex conjugate of its right counterpart $\mathbf{E_r}$ but some boundary conditions, such as Floquet-Bloch pseudo-periodicity conditions, may
 make the explicit resolution of the adjoint problem still necessary.
%
%
%

We can now compute the term of our Keldy\v{s} denominator involving $\TE'$ for any eigentriplets $(\lambda,\mathbf{E_l},\mathbf{E_r})$ where vector fields satisfy the following relations: $\Curl \mathbf{E_r}= - \lambda \, \mathbf{B_r} = - \lambda \, \mu(\lambda) \, \mathbf{H_r} $ and its left counterpart $\Curl \overline{\mathbf{E_l}}= - \lambda \, \overline{\mathbf{B_l}} = - \lambda \, \mu(\lambda)^T \, \overline{\mathbf{H_l}}$.
 We note
 $\TE^{D}(\lambda) \mathbf{E_r} = \Curl \, ( \mu(\lambda)^{-1} \Curl \,   \mathbf{E_r})$
, we use the following formula $(A(z)^{-1})'= -A(z)^{-1} A'(z) A(z)^{-1}$ for the derivation of an inverse matrix depending on a complex parameter, and we have:
\begin{align*}
& \int_\Omega \overline{\mathbf{E_l}} \cdot \left( {\TE^D}'(\lambda) \mathbf{E_r} \right) \, d\Omega  =  \int_\Omega \overline{\mathbf{E_l}} \cdot  \Curl \, ( (\mu(\lambda)^{-1})' \Curl \, \mathbf{E_r})  d\Omega = \\
& -\int_\Omega (\mu(\lambda)^{-T} \Curl \overline{\mathbf{E_l}})  \cdot \left(  \mu(\lambda)' \mu(\lambda)^{-1} \Curl \, \mathbf{E_r} \right ) \, d\Omega + b.t.=\\
& - \int_\Omega (-\lambda  \overline{\mathbf{H_l}})  \cdot \left(  \mu(\lambda)' (-\lambda \mathbf{H_r}) \right) \,   d\Omega + b.t.=\\
& - \int_\Omega  \overline{\mathbf{H_l}}  \cdot  \left( \lambda^2 \, \mu(\lambda)' \, \mathbf{H_r} \right ) \,  d\Omega + b.t.,
\end{align*}
where here
$b.t. = \int_{\partial\Omega}(\overline{\mathbf{E_l}} \times  (\mu(\lambda)^{-1})' \Curl \, \mathbf{E_r})\cdot\mathbf{n} dS$.
Finally, the complete modal expansion formula is:
\begin{equation}
\mathbf{E_s}=  -\sum_{k=1}^{n} \frac{\lambdas}{(\lambdas-\lambda_k)} \frac{\left( \int_{\Omega} \overline{\mathbf{E_l}}_k \cdot \mathbf{j} \,d\Omega \right ) \, \mathbf{E_r}_k}{\int_\Omega \overline{\mathbf{E_l}}_k \cdot \left(  \TE'(\lambda_k) \mathbf{E_r}_k \right ) \,d\Omega}  \label{electric_keldysh}
\end{equation}
with $\int_\Omega \overline{\mathbf{E_l}}_k \cdot \left(  \TE'(\lambda_k) \mathbf{E_r}_k \right ) \,d\Omega =\int_\Omega  \overline{\mathbf{E_l}}_k \cdot \left[ (\lambda_k^2 \varepsilon(\lambda_k))' \, \mathbf{E_r}_k \right ] - \overline{\mathbf{H_l}}_k \cdot \left[ \lambda_k^2 \, \mu(\lambda_k)' \, \mathbf{H_r}_k \right ]  \, d\Omega +
\int_{\partial\Omega}(\overline{\mathbf{E_l}_k} \times  \left( \zeta(\lambda_k)  \mathbf{H_r}_k) \right ) \cdot\mathbf{n} dS $ where $\zeta(\lambda_k) :=  \lambda_k \, \mu(\lambda_k)^{-1}  \, \mu(\lambda_k)'$.
When only the permittivity is dispersive and the boundary conditions allow to discard the boundary terms, a quite common case, $\mathbf{E_l}_k$ is simply equal to $\overline{\mathbf{E_r}}_k$ and the formula becomes:
\begin{equation}
\mathbf{E_s}=  -\sum_{k=1}^{n} \frac{\lambdas}{(\lambdas-\lambda_k)} \frac{\left( \int_{\Omega} \mathbf{E_r}_k \cdot \mathbf{j} \,d\Omega \right ) \, \mathbf{E_r}_k}{\int_\Omega \mathbf{E_r}_k \cdot \left(  (\lambda_k^2 \varepsilon(\lambda_k) )' \mathbf{E_r}_k \right ) \,d\Omega}  \label{electric_keldysh_simple}
\end{equation}
Nevertheless, the full formula will be useful in the setup of dispersive PML for the determination of leaky modes.
\textit{Mutatis mutandis} $( \mathbf{H}\leftrightarrow\mathbf{E},\mu \leftrightarrow \varepsilon)$, the formula based on the magnetic formulation is derived according to exactly the same guidelines.


Since the eigenfields $\mathbf{E_r}_k$ and $\mathbf{E_l}_k$ are both determined up to a non-zero complex factor, it is always possible to choose the amplitude of $\mathbf{E_r}_k$ and $\mathbf{E_l}_k$ in such a way that the denominator in (\ref{electric_keldysh}) is equal to $1$ but there is still one degree of freedom of choice for the amplitude since $\alpha \mathbf{E_r}_k$ and $\mathbf{E_l}_k/ \alpha$, for any $\alpha \neq 0 \in \C$ still gives the same denominator.
It is only in the case where $\mathbf{E_r}_k$ and $\mathbf{E_l}_k$ can be the complex conjugate of each others that the normalization of the denominator can be associated to a normalization of the eigenvectors.
%
%
%
The formulas (\ref{electric_keldysh}) and (\ref{electric_keldysh_simple}) can be directly transposed at the discrete level in any decent Finite Element Modeling software such as  the open source package Onelab/Gmsh/GetDP (\texttt{http://onelab.info})\cite{Dular-GetDP,gmsh}. For all practical purposes, the number $n$ of terms to be considered in (\ref{electric_keldysh}) is related to the required numerical accuracy.

\section{A practical example}
\label{sec:examples}
\begin{figure}[ht]
\centering
\begin{tikzpicture}[scale=0.8]
\def \xL{10};
\def \yL{\xL /2.};
\def \xA{0.5*\xL};\def \yA{0.} ;
\def \xB{0.7*\xL};\def \yB{0.} ;
\def \xC{0.48*\xL};\def \yC{\yL} ;
\def \xD{0.5*\xL};\def \yD{\yL} ;
\def \xM{0.25*\xA+0.25*\xB+0.25*\xC+0.25*\xD};
\def \yM{0.25*\yA+0.25*\yB+0.25*\yC+0.25*\yD};
\def \xS{0.75*\xL}; \def \yS{0.75*\yL};
\def \xP{0.75*\xL}; \def \yP{0.25*\yL};
\coordinate (L) at (\xL,\yL);
\coordinate (A) at (\xA,\yA);
\coordinate (B) at (\xB,\yB);
\coordinate (C) at (\xC,\yC);
\coordinate (D) at (\xD,\yD);
\coordinate (M) at (0.52*\xL,0.4*\yL);
\coordinate (Mu) at (0.3*\xL,0.4*\yL);
\coordinate (Mt) at (0.24*\xL,.8*\yL);
\coordinate (Ms) at (\xL/6.,0.6*\yL);
\coordinate (Orig) at (0,0) ;
\coordinate (S) at (\xS,\yS);
\coordinate (P) at (\xP,\yP);
\filldraw[fill=red!10!white,rotate=0]   (Orig) rectangle (L) ;
\filldraw[fill=green!10!white,rotate=0]  (A) -- (B)--(D)--(C)--cycle ;
\draw (Orig) node[anchor=north] {$O(0,0)$};
\draw (L) node[anchor=south] {$M(L,L/2)$};
\draw (A) node[anchor=north] {$A_1(x_1,0)$};
\draw (B) node[anchor=north] {$A_2(x_2,0)$};
\draw (C) node[anchor=south east] {$A_3(x_3,L/2)$};
\draw (D) node[anchor=south west] {$A_4(x_4,L/2)$};
\fill [red] (S) circle (2pt) ;
\draw (S) node[anchor=south] {$S(x_S,y_S)$};
\fill [black] (P) circle (2pt) ;
\draw (P) node[anchor=south] {$P(x_P,y_P)$};
\draw (Mu) node[anchor=north west] {\Large$\Omega^0$};
\path (Ms) node(a) [rectangle,rotate=0] {$\hat{\chi}(\mathbf{x},\omega)(\mathbf{x})=0$};
\draw (M) node[anchor=north west] {\Large$\Omega_{\mathrm{D}}^1$};
\path (Mt) node(c) [rectangle,rotate=0] {$\hat{\chi}(\mathbf{x},\omega)(\mathbf{x})=\frac{a_1}{\omega-\omega_1^{\varepsilon}} + \frac{a_1}{\omega+\overline{\omega}_1^{\varepsilon}}$};
\draw[thick,red,->] (a) -- (Mu);
\draw[thick,red,->] (c) -- (M);
\end{tikzpicture}
\caption{A box made of a part filled with a dispersive medium $\Omega_D^1$ and the other with a vacuum $\Omega^0$.
The system is entirely determined by the geometrical constants $x_1$, $x_2$, $x_3$, $x_4$ and the position of the source $S(x_S,y_S)$.
In our example, we have $x_1= 0.5 \, L$, $x_2=0.7\,  L$, $x_3= 0.48 \, L$ and $x_4=0.5 \, L$.
As for the position of the source, we have $x_S=0.75 \, L$ and $y_S=0.75\, L/2$.
The field will be later probed at $P$ with $x_P=x_S$ and $y_P=0.25\, L/2$.}
\label{fig:scheme}
\end{figure}
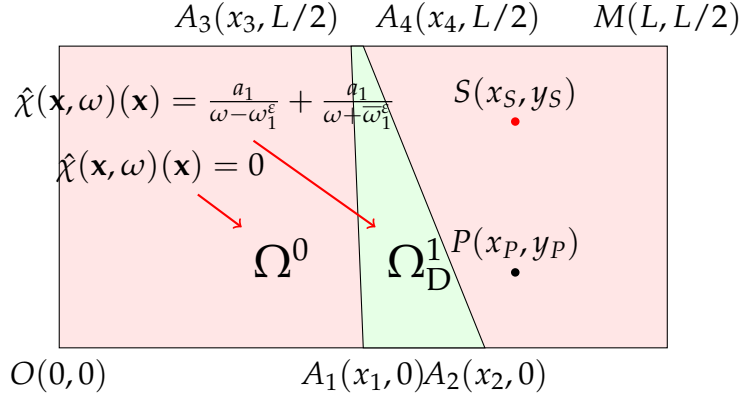
\begin{figure}[ht]
  \centering
  \includegraphics[width=.6\linewidth]{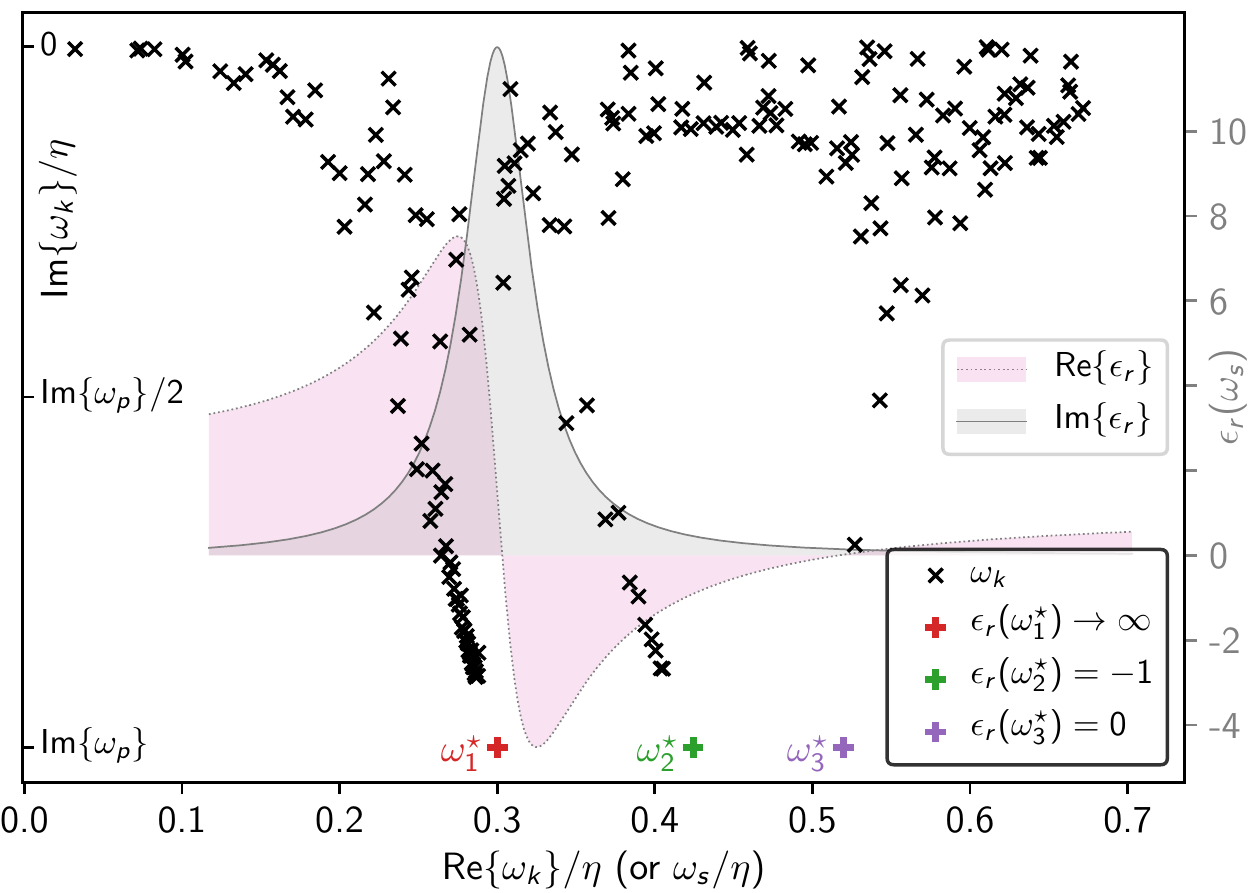}
  \caption{Spectrum of complex resonances (right ordinate axis).
  Relative permittivity for a real frequency (left ordinate axis).}
  \label{fig:spectrum}
\end{figure}

We consider a frequency dispersive inclusion {$\Omega_{\mathrm{D}}^1$} inside a perfectly conducting and empty cavity {$\Omega^0$}
as shown in Fig.~\ref{fig:scheme}. The relative permittivity of the box is a function defined by part inside {$\Omega^0$} as
$\varepsilon_r(\mathbf{x},\omega)=1+\hat{\chi}(\mathbf{x},\omega)$, where the electric susceptibility is chosen as a Lorentz model :
\begin{equation}\label{eq:chi}
  \hat{\chi}(\mathbf{x},\omega)=\frac{a_1(\mathbf{x})}{\omega-\omega_1^{\varepsilon}} + \frac{a_1(\mathbf{x})}{\omega+\overline{\omega}_1^{\varepsilon}}.
\end{equation}
This model can be extended by adding more poles to fit realistic materials \cite{Garcia-Vergara:17}. In the air region
{$\Omega^0$}, $a_1$ is null. In the dispersive region  $\Omega^1_D$, the real part of $\omega_p$ is
chosen in the vicinity of the resonant frequencies of the cavity and its imaginary part small
($\omega_p/\eta=0.3-0.025i$ and $a_1/\eta=-0.3$ with $\eta=4\pi c/L$).
In other words, we are facing here a structure exhibiting both strong interacting material and geometric resonances. The relative permittivity for real frequencies is shown in Fig.~(\ref{fig:spectrum}).

The spectrum of complex resonances computed using Finite Elements
\cite{gmsh,Dular-GetDP,Hernandez:2005:SSF,brule2016calculation,demesy2018eigenmode} is shown in Fig.~\ref{fig:spectrum} (left
ordinate axis). Its general shape is consistent with earlier studies \cite{brule2016calculation} and
its main features are recalled hereafter. At low
frequencies, resonances correspond to the modes of two air boxes surrounding {$\Omega_{\mathrm{D}}^1$}. At
high frequencies, the spectrum of complex resonances tends towards the spectrum of the free Laplacian, as
the permittivity tends towards $\epsilon_0$. The imaginary part of the complex resonances is lower bounded by
$\mathrm{Im}\{\omega_p\}$. Finally, three particular points are emphasized. The first one $\omega^\star_1=\omega_1^{\varepsilon}$
corresponds of course to the pole of the permittivity ($\varepsilon_r(\omega^\star_1)\rightarrow\infty$,
red cross in Fig.~\ref{fig:spectrum}). It forms an accumulation
points for the resonances with corresponding eigenvectors exhibiting spatial frequencies within
{$\Omega_{\mathrm{D}}^1$} tending to zero. The second one correspond to the plasmon branch where
$\varepsilon_r(\omega^\star_2)=-1$ (green cross in Fig.~\ref{fig:spectrum}). It is another an accumulation point
for the \textit{plasmonic resonances} with corresponding eigenvectors oscillating with spatial frequencies tending
to zero on the two interfaces between {$\Omega_{\mathrm{D}}^1$} and {$\Omega^0$}. The last one is the
zero of the permittivity ($\varepsilon_r(\omega^{\star}_3)=0$, purple cross in Fig.~\ref{fig:spectrum}).

The flexibility of the non-linear eigenvalue solver of the SLEPc library \cite{Hernandez:2005:SSF} allows to
define a rectangular target zone to stay away the three particular points (sharing the same imaginary part in the present case). Only eigenvalues with imaginary parts greater than $0.8\,\textrm{Im}\{\omega_p\}$ are selected. Closer to the accumulation points, the corresponding eigenvectors oscillate faster than the mesh size and become  spurious.

\begin{figure*}[ht]
  \includegraphics[width=1.\textwidth]{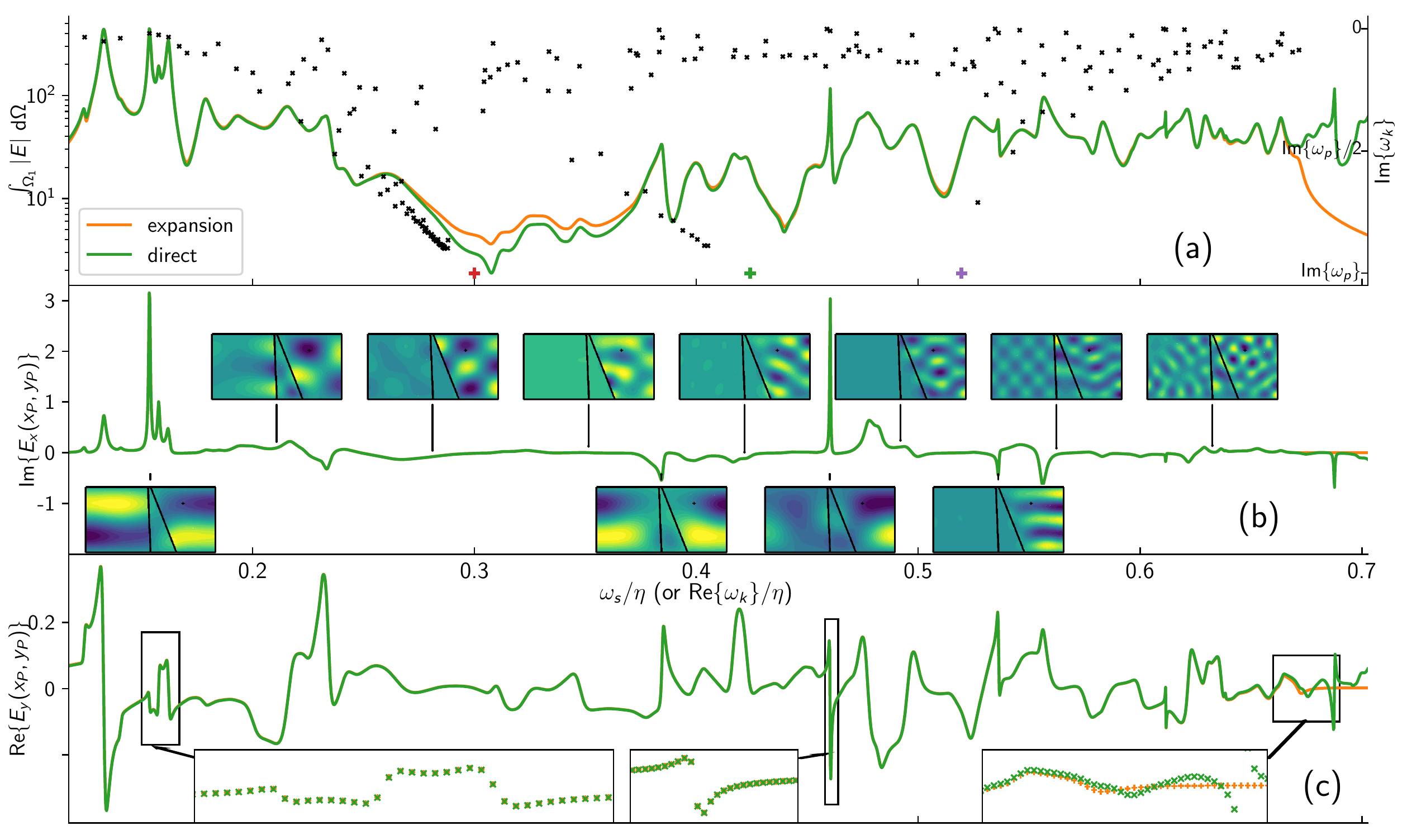}
  \caption{Total field obtained by expansion (orange solid curves) or by solving a direct
  problem classically (green solid curves).
  (a) Integral over $\Omega_D^1$ of the norm of the total electric field (left axis) and spectrum of
  the complex resonances (right axis, same data as in Fig.~\ref{fig:spectrum}).
  (b) Imaginary part of the $x$ component of the electric fields at ($x_p,y_p$).
  Each inset is the map of Im$\{E_x\}$ where the min value is in dark blue and the max value in yellow.
  (c) Real parts of the $y$ component of the electric fields at ($x_p,y_p$). Note
  that the real part is singular on the source point itself for the direct problems. Each inset is a zoom
  on the calculated points (same color convention as above).
  }
  \label{fig:fields}
\end{figure*}

Now, let us
(i) consider a simple oriented delta as a source,
(ii) solve, for a real frequency
     $\omegas$, the direct classical scattering problem
     $-\Curl \, \Curl \, \mathbf{E} + (\omegas/c)^2 \varepsilon_r(\omegas) \, \mathbf{E} =\delta(x_S,y_S)\mathbf{x}$, and
(iii) compare the total field obtained (green lines in Fig.~\ref{fig:fields}) to the one obtained using the reconstruction
formula in Eq.~\ref{electric_keldysh_simple} (yellow lines, with the 200 modes obtained previously).
This is a part of the 2D dyadic Green function of the structure. The results for the mean value of the electric field inside $\Omega_D^1$
($\int_{\Omega^\mathrm{D}_1}|\mathbf{E}|\,\mathrm{d}\Omega$ represented in logscale in Fig.~\ref{fig:fields}(a)), for the
imaginary part of the field at point $P$ (Im$\{E_x(x_P,y_P)\}$ in Fig.~\ref{fig:fields}(b)) and for the real part of the
part of the field at point $P$ (Re$\{E_x(x_P,y_P)\}$ in Fig.~\ref{fig:fields}(c)) show remarkable agreement.
The greatest discrepancies, only noticeable in logscale in Fig.~\ref{fig:fields}(a), are observed
for $\omegas$ near Re$\{\omega_p\}$ and near the plasmonic accumulation point. Let us recall here that only the
normalized eigenvalues with imaginary parts greater than $0.8 \, \mathrm{Im}\{\omega_p\}$ are taken into account in the expansion.
Note that agreement collapses after
$\omegas \gtrsim 0.664\eta$ 
because the real part of the last computed eigenvalue is $0.671\eta$.
The $L^2$ relative  error between the expansion and the direct problem on the normalized frequency range
$\omegas \in[0.12,0.66]$ is
$1.63\%$ on the quantity $\int_{\Omega^\mathrm{D}_1}|\mathbf{E}|\,\mathrm{d}\Omega$,
$0.12\%$  on the quantity Im$\{E_x(x_P,y_P)\}$, and
$1.11\%$  on the quantity Re$\{E_y(x_P,y_P)\}$. Note that this last quantity is singular on the source point
and still reconstructed with high accuracy at point $P$. The imaginary part of the Green function, most quantity useful describing the for instance the resonant dipole-dipole interaction, is regular and reconstructed everywhere in $\Omega^0$ with a relative error of less than $0.5\%$.
\section{Conclusion}
We have shown on a very simple case (albeit very strongly dispersive) that the modal expansion shows a remarkable accuracy on a wide frequency range. All the modes are used in the expansion at all the excitation frequencies but only a very small number with the real part of the eigenfrequency close to the excitation frequency are really contributing. The most important further development of this work will be the modal expansion for open structures involving leaky quasi-normal modes (QNM)\cite{vial2014quasimodal}. In the Finite Element case, this requires the use of Perfectly Matched Layers (PML) and the use of dispersive PML would allow the treatment of a wide range of frequencies. In this case, some dispersive permeability will be introduced and the full formula (\ref{electric_keldysh}) will be necessary. The work will also be extended to a wide range of structures including diffraction grating and 3D nanoresonators. The modal expansion will also be used in time domain \textit{e.g.} for the fast computation of the response of a dispersive scatterer to an incident pulse.
\section*{Acknowledgement}
This research was supported by ANR RESONANCE project, grant ANR-16-CE24-0013 of the French Agence Nationale de la Recherche.

\bigskip
\bibliographystyle{plain}
\bibliography{sample}


\end{document}